\documentclass[aps,showpacs, nofootinbib]{revtex4}
\newcommand{\be}{\begin{equation}}
\newcommand{\ee}{\end{equation}}
\newcommand{\ben}{\begin{eqnarray}}
\newcommand{\een}{\end{eqnarray}}

\newcommand{\cL}{{\cal L}}
\newcommand{\cH}{{\cal H}}

\newcommand{\p}{\partial}
\newcommand{\na}{\nabla}

\newcommand{\ep}{\epsilon}
\newcommand{\bep}{\bar \epsilon}

\newcommand{\bdel}{\bar \delta}


\begin{document}

\title{First Law of Black Saturn Thermodynamics}
%%%%%%%%%%%%%%%%%%%%%%%%%%%%%%%%%%%%%%%%%%%%%%%%%%%%%%%%%%%%%%

\author{Marek Rogatko}
\affiliation{Institute of Physics \protect \\
Maria Curie-Sklodowska University \protect \\
20-031 Lublin, pl.~Marii Curie-Sklodowskiej 1, Poland \protect \\
rogat@tytan.umcs.lublin.pl \protect \\
rogat@kft.umcs.lublin.pl}

%%%%%%%%%%%%%%%%%%%%%%%%%%%%%%%%%%%%%%%%%%%%%%%%%%%%%%%%%%%%%%%%%%%%
\date{\today}
%\pacs{04.30.Nk, 04.40.-b}

%%%%%%%%%%%%%%%%%%%%%%%%%%%%%%%%%%%%%%%%%%%%%%%%%%%%%%%%%%%%%%%%%%%%%%%%%%%%%%%%%%%%%%%%%%%%%%%%%%%
\begin{abstract}
The {\it physical process} version and {\it equilibrium state} version of the first law
of thermodynamics for a black object consinsting of $n$-dimensional charged stationary axisymmetric
black hole surrounded by $a$ black rings, the so-called black Saturn was derived. 
The general setting for our derivations is $n$-dimensional dilaton gravity with $p + 1$ strength form fields.

\end{abstract}
%%%%%%%%%%%%%%%%%%%%%%%%%%%%%%%%%%%%%%%%%%%%%%%%%%%%%%%%%%%%%%%%%%%%%%%%%%%%%%%%%%%%%%%%%%%%%%%%%%%%

\maketitle

%%%%%%%%%%%%%%%%%%%%%%%%%%%%%%%%%%%%%%%%%%%%%%%%%%%%%%%%%%%%%%%%%%%%%%%%%%%%%%%%%%%%%%%%%%%%%%%%%%%%%
\section{Introduction}
The idea that spacetime may have been more than four-dimensional manifold acquired popularity recently. 
One of the most promising approaches to unification of fundamental interactions of Nature is
superstring/ M-theory. Those unified theories are formulated in the spacetime of higher dimensions.
Consequently, this approach triggers continuously growing interests in studying properties of black
holes in higher dimensional theories. It happened that they reveal a variety of new and interesting properties.
The uniqueness theorem for 
static $n$-dimensional black holes was quite well established \cite{uniq}.
But for stationary axisymmetric $n$-dimensional solutions
the situation is far from obvious. In fact, it was shown \cite{emp02} that even in five-dimensional
spacetime a new type of black object emerged, the so-called {\it black ring}. This solution has $S^2 \times S^1$
topology of the event horizon and is equipped in the same mass and angular momentum as a spherical 
five-dimensional stationary axisymmetric black hole. However, if one assumes the topology of
black hole event horizon as $S^3$ the uniqueness
proof can be established (see Ref.\cite{mor04} for the vacuum case and Ref.\cite{rog04a} for the 
stationary axisymmetric self-gravitating $\sigma$-model).
For a review of a black ring story see \cite{emp06} and references therein.
\par
But it turned out that we can have more complicated black object, not only black hole or black ring.
In Ref.\cite{elv07} by means of the inverse scattering method an exact asymptotically 
flat five-dimensional solution describing black Saturn, i.e.,
a spherical black hole surrounded by a black ring, was derived. It was also revealed \cite{elv07a}
that the configurations that approach maximal entropy in five-dimensional asymptotically flat 
vacuum gravity for fixed mass and angular momentum are black Saturns. Because of the growing interests
in such black object our main aim will be to study the first law of mechanics for it. In our paper
we shall look for the {\it physical process} version of the first law of black Saturn thermodynamics
as well as {\it equilibrium state} version of it.
\par
The {\it physical process} version of the first law of black object thermodynamics
is realized by changing a stationary black hole or black ring by some infinitesimal 
physical process, e.g., by throwing matter into black object. If we assume that the 
final state of black object settles down to a 
stationary one, we can extract the changes of black object's parameters and 
in this way obtain information about the first law of its mechanics. 
The {\it physical process } version of the first 
law of black hole thermodynamics was extensively studied in the context of Einstein and Einstein-Maxwell (EM)
theory in Refs.\cite{wal94,gao01} and
in
Einstein-Maxwell axion-dilaton (EMAD) gravity being the low-energy limit of the heterotic
string theory in Ref.\cite{rog02}. While the case  Einstein gravity coupled to
$(n-2)$-gauge form field strength was treated in Ref.\cite{rog05}.
The case of black rings in higher dimensional dilaton gravity containing
$(p+1)$-form field strength was considered in \cite{rog05br}, being
the simplest generalization of five-dimensional one in which stationary black ring
solution has been provided \cite{elv05}.
\par
The other attitude to the problem of the first law of black hole thermodynamics is the so-called
{\it equilibrium state} version. It was studied in the seminal paper of
Bardeen, Carter and Hawking \cite{bar73}. This attitude is based on taking into account
the linear perturbations of a stationary electrovac black hole to another one.
In Ref.\cite{sud92}
arbitrary asymptotically flat perturbations of a stationary 
black hole were considered, while the first law of black hole thermodynamics valid for
an arbitrary diffeomorphism invariant Lagrangian 
with metric and matter fields possessing stationary and axisymmetric
black hole solutions were given in Refs.\cite{wal93}-\cite{iye97}. The cases of higher 
curvature terms and higher derivative terms in the metric
were considered in
\cite{jac}, while the situation when the Lagrangian is an arbitrary function of metric,
Ricci tensor and a scalar field was elaborated in Ref.\cite{kog98}.
In Ref.\cite{gao03}
of a charged rotating black hole
where fields were not smooth through the event horizon was treated.
In Ref.\cite{cop05}, the authors using the notion of
bifurcate Killing horizons and taking into account dipole charges were managed
to find the first law of black hole thermodynamics for black ring solutions.
In the higher dimensional gravity containing $(p + 1)$-form field strength and dilaton fields
the first law of black ring mechanics choosing an arbitrary cross
section of the event horizon to the future of the bifurcation surface was derived in Ref.\cite{rog05br1}.
On the other hand,
the {\it physical process} version and the {\it equilibrium state} version of the first law
of black ring thermodynamics in $n$-dimensional Einstein gravity with Chern-Simons term
were derived in Ref\cite{rog07}. Using the covariant cohomological methods to the conserved charges
for $p$-form gauge fields coupled to gravity the first law of thermodynamics was found in Ref.\cite{com07}.
\par
In our paper we shall
consider the first law of black Saturn thermodynamics in generalized $n$-dimensional dilaton
gravity with $p + 1$ strength forms being
the simplest generalization of a five-dimensional theory which
admits stationary black hole and black ring solutions. 
The model under considerations will be composed of axisymmetric stationary $n$-dimensional black hole
surrounded by $a$ black rings. Sec.II will be devoted to the {\it physical process} version of the first law of 
black Saturn thermodynamics. In
Sec.III we shall consider the {\it equilibrium state} version of the first law
of the black objects by choosing the arbitrary cross sections of event horizons of the 
adequate components of black Saturn to the future of their bifurcation surfaces. 
This method allows one to treat fields which are not necessary 
smooth through the adequate event horizons. In Sec.IV we concluded our investigations.
%%%%%%%%%%%%%%%%%%%%%%%%%%%%%%%%%%%%%%%%%%%%%%%%%%%%%%%%%%%%%%%%%%%%%%%%%%%%%%%%%%%%
%%%%%%%%%%%%%%%%%%%%%%%%%%%%%%%%%%%%%%%%%%%%%%%%%%%%%%%%%%%%%%%%%%%%%%%%%%%%%%
\section{Physical process version of the first law of black Saturn mechanics}
The general setting for our theory will be $n$-dimensional dilaton gravity with
$p + 1$-form field strength. This theory is the simplest generalization of the five-dimensional 
one with three form field strength and dilaton fields which contains stationary 
black hole and black ring solution. The Lagrangian for the theory under consideration is given by
\be
{\bf L } = {\bf \ep} \bigg(
{}^{(n)}R - {1 \over 2} \na_{\mu}\varphi \na^{\mu} \varphi - {1\over 2 (p + 1)!} e^{-{\alpha} \varphi}
H_{\mu_{1} \dots \mu_{p+1}} H^{\mu_{1} \dots \mu_{p+1}}
\bigg),
\label{lag}
\ee
where by $ {\bf \ep}$ we denote the volume element,
$\phi$ is the dilaton field while
$H_{\mu_{1} \dots \mu_{p+1}} = (p + 1)! \na_{[ \mu_{1}} B_{{\mu_{2} \dots \mu_{p+1}]}} $ is $(p + 1)$-form field strength.
Let us assume that in the theory described by the relation (\ref{lag}) the black Saturn solution is given.
By the notion black Saturn we shall understand $n$-dimensional axisymmetric stationary black hole surrounded by
$a$ black rings. We begin with the {\it physical version} of the first law of black Saturn thermodynamics.
In order to find this law one ought to get the explicit variations of mass and angular momenta. Evaluating
the variations of the adequate fields, we achieve the following:
\ben \label{dl}
\delta {\bf L} &=& {\bf \epsilon} \bigg(
G_{\mu \nu} - T_{\mu \nu}(B, \varphi) \bigg)~ \delta g^{\mu \nu}
- {\bf \epsilon} \na_{j_{1}} \bigg( e^{-{\alpha} \varphi}H^{j_{1} \dots j_{p+1}} \bigg) 
\delta B_{j_{2} \dots j_{p+1}}  \\ \nonumber
&+& {\bf \epsilon} \bigg( \na_{\mu}\na^{\mu} \varphi + {\alpha \over 2 (p + 1)!} e^{-{\alpha} \varphi}
H_{\mu_{1} \dots \mu_{p+1}} H^{\mu_{1} \dots \mu_{p+1}} \bigg)~\delta \varphi
+ d {\bf \Theta},
\een
where the energy momentum for the considered setting yields
\ben
T_{\mu \nu}(B, \varphi) &=&
{1 \over 2} \na_{\mu} \varphi \na_{\nu} \varphi - {1 \over 4} g_{\mu \nu} \na_{\alpha} \varphi
\na^{\alpha} \varphi \\ \nonumber
 &+& {1\over 2 (p + 1)!} e^{-{\alpha} \varphi}
\bigg[ (p + 1) H_{\mu \nu_{2} \dots \nu_{p+1}} H_{\nu}{}{}^{\nu_{2} \dots \nu_{p+1}}
- {1 \over 2} g_{\mu \nu} H_{\mu_{1} \dots \mu_{p+1}} H^{\mu_{1} \dots \mu_{p+1}}
\bigg].
\een
Having in mind the relation (\ref{dl}) one achieves
the symplectic $(n - 1)$-form
$\Theta_{j_{1} \dots j_{n-1}}[\eta_{\alpha}, \delta \eta_{\alpha}]$ in the following form:
\be
\Theta_{j_{1} \dots j_{n-1}}[\eta_{\alpha}, \delta \eta_{\alpha}] =
\ep_{\mu j_{1} \dots j_{n-1}} \bigg[
\omega^{\mu} - e^{-{\alpha} \varphi} H_{m \nu_{2} \dots \nu_{p+1}}~\delta B_{\nu_{2} \dots \nu_{p+1}}
- \na^{m} \varphi~ \delta \varphi \bigg],
\ee 
where $\omega_{\mu} = \na^{\alpha} \delta g_{\alpha \mu} - \na_{\mu} 
\delta g_{\beta}{}{}^{\beta}$ and $\eta_{\alpha}$ stands for the fields in the underlying theory
while their variations are denoted by $\delta \eta_{\alpha}$.
\par 
As in Refs.\cite{gao01,rog07}, we identify variations of fields with a general coordinate transformations
induced by an arbitrary Killing vector field $\xi_{\alpha}$. In the next step, we calculate
the Noether $(n - 1)$-form with respect to this above mentioned Killing vector, i.e., 
${\cal J}_{j_{1} \dots j_{n-1}} = \ep_{m j_{1} \dots j_{n-1}} {\cal J}^{m}
\big[\eta_{\alpha}, {\cal L}_{\xi} \eta_{\alpha}\big]$, namely
\ben
{\cal J}_{j_{1} \dots j_{n-1}} &=& 
d \bigg( Q^{GR} + Q^B \bigg)_{j_{1} \dots j_{n-1}}
+ 2 \ep_{m j_{1} \dots j_{n-1}} \bigg( G^{m}{}{}_{\eta} - T^{m}{}{}_{\eta}(B, \varphi)
\bigg) \xi^{\eta} \\ \nonumber
&+& p~\ep_{m j_{1} \dots j_{n-1}}~ \xi^{d} B_{d \alpha_{3} \dots \alpha_{p+1}}~
\na_{\alpha_{2}} \bigg( e^{-{\alpha} \varphi} H^{m \alpha_{2} \dots \alpha_{p+1}} \bigg),
\een
where $Q_{j_{1} \dots j_{n-2}}^{GR}$ yields
\be
Q_{j_{1} \dots j_{n-2}}^{GR} = - \ep_{j_{1} \dots j_{n-2} a b} \na^{a} \xi^{b},
\ee
while $Q_{j_{1} \dots j_{n-2}}^{B}$ has the following form:
\be
Q_{j_{1} \dots j_{n-2}}^{B} = {p \over (p + 1)!} \ep_{m \alpha j_{1} \dots j_{n-2}}
~\xi^{d}~B_{d \alpha_{3} \dots \alpha_{p+1}}~ e^{-{\alpha} \varphi} H^{m \alpha \alpha_{3} \dots \alpha_{p+1}}.
\ee
$Q_{j_{1} \dots j_{n-1}} = (Q^{GR} + Q^{B})_{j_{1} \dots j_{n-1}}$ constitutes the Noether charge for the
$n$-dimensional dilaton gravity with $p + 1$-form strength fields. One has in mind that \cite{gao01}
${\cal J}[\xi] = dQ[\xi] + \xi^{\alpha} {\bf C}_{\alpha}$,
where ${\bf C}_{\alpha}$ is an $(n-1)$-form constructed from dynamical fields, i.e., from
$g_{\mu \nu}$, $(p + 1)$-form field $H^{j_{1} \dots j_{p+1}}$ and dilaton fields. 
${\bf C}_{\alpha}$ reduces to the following form:
\be
C_{d j_{1} \dots j_{n-1}} = 2 \ep_{m j_{1} \dots j_{n-1}}
\bigg[ G_{d}{}{}^{m} - T_{d}{}{}^{m}(B, \varphi) \bigg] +
p~ \ep_{m j_{1} \dots j_{n-1}}~\na_{\alpha_{2}} \bigg( 
e^{-{\alpha} \varphi} H^{m \alpha_{2} \dots \alpha_{p+1}} \bigg)
B_{d \alpha_{3} \dots \alpha_{p+1}}.
\ee
When ${\bf C}_{\alpha} = 0$ one gets the source-free Eqs. of motion but on the other hand,
we get the following:
\ben
G_{\mu \nu} - T_{\mu \nu}(B, \varphi) &=& T_{\mu \nu}(matter) , \\
\na_{\mu_{1}} \bigg( e^{-{\alpha} \varphi} H^{\mu_{1} \dots \mu_{p+1}} \bigg)
 &=& j^{\mu_{2} \dots \mu_{p+1}}(matter).
\een
Let us assume further that $(g_{\mu \nu}, B_{\alpha_{1} \dots \alpha_{p}}, \varphi)$ are solutions 
of source-free equations of motion and 
$(\delta g_{\mu \nu},~\delta B^{\alpha_{1} \dots \alpha_{p}},~\delta \varphi)$
are the linearized perturbations satisfying Eqs. of motion with sources
$\delta T_{\mu \nu}(matter)$ and $ \delta j^{\mu_{1} \dots \mu_{p}}(matter)$, then we reach to the
relation of the form as follows:
\be
\delta  C_{a j_{1} \dots j_{n-1}} = 2 \ep_{m j_{1} \dots j_{n-1}}
\bigg[ \delta T_{a}{}{}^{m}(matter) + p~B_{a \alpha_{3} \dots \alpha_{p+1}}~ 
\delta j ^{m \alpha_{3} \dots \alpha_{p+1}}(matter) \bigg].
\ee
The fact that the Killing vector field $\xi_{\alpha}$ describes also a symmetry of the background
matter field, it provides the formula for a conserved quantity connected with $\xi_{\alpha}$. It follows directly
\ben \label{hh}
\delta H_{\xi} &=& - 2 \int_{\Sigma}\ep_{m j_{1} \dots j_{n-1}} \bigg[
\delta T_{a}{}{}^{m}(matter) \xi^{a} + 
 p~\xi^{a}~B_{a \alpha_{3} \dots \alpha_{p+1}}~ 
\delta j ^{m \alpha_{3} \dots \alpha_{p+1}}(matter) \bigg]
 \\ \nonumber
&+& \int_{\p \Sigma}\bigg[
\delta Q(\xi) - \xi \cdot \Theta \bigg].
\een
As in the case of black hole or black ring first law of mechanics,
let us 
choose $\xi^{\alpha}_{(i)}$ to be an asymptotic time translation $t^{\alpha}$ for each black object.
Just we draw the conclusion 
that the variation of the ADM mass for each object will be given as follows:
\ben \label{mm}
\alpha~ \delta M_{(i)} &=& - 2 \int_{\Sigma_{(i)}} \ep_{m j_{1} \dots j_{n-1}} \bigg[
\delta T_{a}{}{}^{m}(matter) t^{a} + p~ t^{a} B_{a \alpha_{3} \dots \alpha_{p+1}}~ 
\delta j ^{m \alpha_{3} \dots \alpha_{p+1}}(matter) \bigg] \\ \nonumber
&+& \int_{\p \Sigma_{(i)}}\bigg[
\delta Q(t) - t \cdot \Theta \bigg],
\een
where $\alpha = {n-3 \over n-2}$ and $i$ stands for the case of black hole or black ring.
Further on, we take into account the Killing vector fields $\psi_{(i)}$ which are responsible
for the rotation in the adequate directions for black hole, and 
$\phi_{(i)}$ which are connected with the rotation of black rings.
It provides finally 
the following relations for angular 
momenta respectively for black hole and black rings. The variations of angular momenta for
$n$-dimensional black hole imply
\ben \label{jb}
\delta J_{(i)}^{BH} &=& 2 \int_{\Sigma_{(i)}} \ep_{m j_{1} \dots j_{n-1}} \bigg[
\delta T_{a}{}{}^{m}(matter) \psi_{(i)}^{a} + p~ \psi_{(i)}^{a} B_{a \alpha_{3} \dots \alpha_{p+1}}~ 
\delta j ^{m \alpha_{3} \dots \alpha_{p+1}}(matter) \bigg] \\ \nonumber
&-& \int_{{\p \Sigma}_{(i)}}\bigg[
\delta Q({\psi}_{(i)}) - {\psi}_{(i)} \cdot \Theta \bigg],
\een
while the variations of angular momenta for surrounding black rings yield
\ben \label{jr}
\delta J_{(i)}^{BR} &=& 2 \int_{{\Sigma}_{(i)}} \ep_{m j_{1} \dots j_{n-1}} \bigg[
\delta T_{a}{}{}^{m}(matter) \phi_{(i)}^{a} + p~ \phi_{(i)}^{a} B_{a \alpha_{3} \dots \alpha_{p+1}}~ 
\delta j ^{m \alpha_{3} \dots \alpha_{p+1}}(matter) \bigg] \\ \nonumber
&-& \int_{{\p \Sigma}_{(i)}}\bigg[
\delta Q({\phi}_{(i)}) - {\phi}_{(i)} \cdot \Theta \bigg].
\een
To consider the {\it physical process} version of the first law of black Saturn thermodynamics,
let us assume that $(g_{\mu \nu},~B_{\alpha_{1} \dots \alpha_{p}},~\varphi)$ are solutions to the source free
Einstein equations with $(p+1)$ form fields and scalar dilaton fields.
Moreover, let $\xi^{\alpha}_{(BH)}$ denotes
the event horizon Killing vector field connected with black hole
\be
\xi^{\mu}_{(BH)} = t^{\mu} + \sum_{i} \Omega_{(i)} \psi^{\mu (i)},
\label{kil}
\ee
and $\xi^{\alpha}_{(BR~(i))}$ denotes the event horizon Killing vector field for $i-th$ black ring
\be
\xi^{\mu}_{(BR~(i))} = t^{\mu} + \sum_{m} \omega_{(m)~(i)} \phi^{\mu (m)}_{(i)},
\label{kilbr}
\ee
Let us perturb the black Saturn by dropping into it some matter. Furthermore,
suppose
that the black Saturn will be not destroyed during this process and it settles down to a stationary
solution \cite{gao01}. 
Our next task will be to find changes of masses and angular momenta of the 
black objects under consideration
according to relations
(\ref{mm})-(\ref{jr}). Changes of the horizons' area will be computed using
the $n$-dimensional Raychaudhuri equation.
In addition, we shall assume that
$\Sigma_{0 (i)}$ is an asymptotically flat
hypersurface which terminating on the $i-th$ event horizon of the considered black objects, i.e.,
black hole or on $a-th$ black ring surrounded the higher dimensional black hole under consideration.
Then, one takes into account 
the initial data on $\Sigma_{0 (i)}$
for a linearized perturbations of
$(\delta g_{\mu \nu},~ \delta B_{\alpha_{1} \dots \alpha_{p}}, \delta \varphi)$
with $\delta T_{\mu \nu}(matter)$ and $\delta j^{\alpha_{2} \dots \alpha_{p+1}}(matter)$. 
We require that $\delta T_{\mu \nu}(matter)$ and $\delta j^{\alpha_{2} \dots \alpha_{p+1}}(matter)$
disappear
at infinity and the initial data for 
$(\delta g_{\mu \nu},~ \delta B_{\alpha_{1} \dots \alpha_{p}}, \delta \varphi)$
vanish in the vicinity of each black object horizon ${\cal H}_{(BH)}$ and ${\cal H}_{(BR(i))}$ on 
the adequate
hypersurface $\Sigma_{0 (i)}$.
The above conditions provide
that for the initial time
$\Sigma_{0 (i)}$, each of the black object is unperturbed. On its own, it provides
the perturbations vanish near the internal boundary $\p \Sigma_{0 (i)}$,
one gets from relations (\ref{mm}) and (\ref{jr}) that the following is fulfilled:
\ben \label{ppp}
\alpha~ \bigg(
\delta M_{BH} &+& \sum_{a} \delta {M_{BR}}^ {(a)}
 \bigg)
-  \sum_{i} \Omega_{(i)} \delta J^{(i)}_{BH} 
- \sum_{a}~ \sum_{i} \omega_{(i) (a)} \delta {J^{(i)}_{BR}}_{(a)} 
= \\ \nonumber
&-& 2 \int_{\Sigma_{0 (BH)}} \ep_{m j_{1} \dots j_{n-1}} \bigg[
\delta T_{k}{}{}^{m}(matter)~ \xi^{k}_{(BH)} + p~ \xi^{k}_{(BH)}~ B_{k \alpha_{3} \dots \alpha_{p+1}}~ 
\delta j ^{m \alpha_{3} \dots \alpha_{p+1}}(matter) \bigg]
\\ \nonumber
&-& 2 \sum_{a}~
\int_{\Sigma_{0(a)}} \ep_{m j_{1} \dots j_{n-1}} \bigg[
\delta T_{k}{}{}^{m}(matter)~ \xi^{k}_{(BR~(a))} + p~ \xi^{k}_{(BR~(a))}~ B_{k \alpha_{3} \dots \alpha_{p+1}}~ 
\delta j ^{m \alpha_{3} \dots \alpha_{p+1}}(matter) \bigg]
\\ \nonumber
&=& \int_{{\cH}_{(BH)}} \gamma ^{\alpha}_{BH}~k_{\alpha (BH)}~\bep_{j_{1} \dots j_{n-1} (BH)}
+ 
\sum_{a}~\int_{{\cH}_{(BR(a))}} \gamma ^{\alpha}_{BR~(a)}~k_{\alpha (BR_{(a)})}~\bep_{j_{1} \dots j_{n-1} (BR (a))}
\een
where $\bep_{j_{1} \dots j_{n-1} (i)} = n^{\delta}_{(i)}~\ep_{\delta j_{1} \dots j_{n-1}}$ while
$n^{\delta}_{(i)}$ is a future directed unit normal to the adequate hypersurface $\Sigma_{0 (i)}$ for each black object.
On the other hand, each
$k_{\alpha (i)}$ is tangent vector to the affinely parametrized null geodesics generators of the adequate
black object event horizon.
Due to the fact of the conservation of each current $\gamma^{\alpha}_{(i)}$ and the assumption
that all of the matter falls into each of the considered black object we replace in Eq.(\ref{ppp})
$n^{\delta}_{(i)}$ by the vector $k^{\delta}_{(i)}$.
Let us recall relation which will be useful in finding
the integrals over the event horizons of the adequate black object
\ben \label{cf}
p!~\cL_{\xi} B_{\alpha_{2} \dots \alpha_{p+1}}~\delta j^{\alpha_{2} \dots \alpha_{p+1}}
&-& \xi^{d}~H_{d \alpha_{2} \dots \alpha_{p+1}}~\delta j^{\alpha_{2} \dots \alpha_{p+1}} \\ \nonumber
&=& p~ p! \na_{\alpha_{2}} \bigg(
\xi^{d}~B_{d \alpha_{2} \dots \alpha_{p+1}}
\bigg)~\delta j^{\alpha_{2} \dots \alpha_{p+1}}.
\een
Having in mind that in the stationary background $\theta$ expansion and
$\sigma_{ij}$ shear vanish and using  $n$-dimensional Raychauduri 
lead us to the fact that
$R_{\alpha \beta} k^{\alpha}_{(i)} k^{\beta}_{(i)} \mid_{{\cH}_{(i)}} = 0$. This in turn implies the following
for each black object:
\be
{1 \over 2}k^{\mu}_{(i)} \na_{\mu} \varphi~ k^{\nu}_{(i)} \na_{\nu} \varphi +
{1 \over 2 p!} e^{- \alpha \varphi}
H_{\mu \mu_{2} \dots \mu_{p+1}} H_{\nu}{}{}^{ \mu_{2} \dots \mu_{p+1}} k^{\mu}_{(i)} k^{\nu}_{(i)} \mid_{{\cH}_{(i)}} = 0.
\ee
Using the fact that $\cL_{k} \varphi = 0$,
it is easily seen that,
$H_{\nu}{}{}^{ \mu_{2} \dots \mu_{p+1}} k^{\nu}_{(i)} = 0$. Because of the fact that 
$H_{\mu \mu_{2} \dots \mu_{p+1}} k^{\mu}_{(i)} k^{\mu_{2}}_{(i)} = 0$, then by asymmetry of $H_{ \mu_{1} \dots \mu_{p+1}}$
it follows that $H_{\mu \mu_{2} \dots \mu_{p+1}} k^{\mu}_{(i)} \sim k_{\mu_{2} (i)} \dots k_{\mu_{p+1} (i)}$.
The pull-back of $H_{\mu}{}{}^{ \mu_{2} \dots \mu_{p+1}} k^{\mu}_{(i)}$ to the adequate event horizon is equal to zero.
Thus,
$\xi^{d (BH)}~H_{d \alpha_{2} \dots \alpha_{p+1}}$ is a closed
$p$-form on the event horizon of the considered black hole.
\par
For the case of $n$-dimensional stationary axisymmetric black hole we get the following:
\ben
&-& 2 \int_{{\cH}_{(BH)}} \ep_{m j_{1} \dots j_{n-1}}
 p~ \xi^{k}_{(BH)}~ B_{k \alpha_{3} \dots \alpha_{p+1}}~\delta j ^{m \alpha_{3} \dots \alpha_{p+1}}(matter) =
\\ \nonumber
&-& \Phi_{(BH)} \int_{{\cH}_{(BH)}} 
\ep_{m j_{1} \dots j_{n-1}}
2 p~\delta j ^{m \alpha_{3} \dots \alpha_{p+1}}(matter) 
= \Phi_{(BH)}~\delta Q_{(BH)},
\een
where $\delta Q_{(BH)}$ denotes the net flux of charge flowing into the considered black hole.
\par
For the case of black rings surrounded the black hole the situation is a little bit more complicated. Namely,
due to the Hodge theorem (see e.g., \cite{wes81}) it may be rewritten
as a sum of an exact and harmonic form. An exact one does not contribute to the above expression
because of the fact that equations of motion must be satisfied.
The only contribution originates from the harmonic part of
$\xi^{d}_{(BR (a))}~H_{d \alpha_{2} \dots \alpha_{p+1}}$. The duality between homology
and cohomology provides that 
there is a harmonic form $\eta$ dual to $n - p- 1$ cycle $\cal S$ in the sense of the equality
of the adequate surface integrals. It follows that the surface terms will have the form of
$\Phi_{(BR(a))}~\delta q_{(BR(a))}$, where $\Phi_{(BR(a))}$ is the constant relating to the harmonic part
of $\xi^{d}_{(BR (a))}~H_{d \alpha_{2} \dots \alpha_{p+1}}$ and $\delta q_{(BR(a))}$ is the variation of a local charge \cite{cop05}.
Summing it all up, we conclude that the following is fulfilled:
\ben \label{rh}
\alpha~ 
\bigg( 
\delta M_{BH} &+& \sum_{a} \delta {M_{BR}}^ {(a)}
 \bigg)
-  \sum_{i} \Omega_{(i)} \delta J^{(i)}_{BH} 
- \sum_{a}~ \sum_{i} \omega_{(i) (a)} \delta {J^{(i)}_{BR}}_{(a)} = \\ \nonumber
&+&
\Phi_{(BH)}~\delta Q_{(BH)} + \sum_{a} \Phi_{(BR (a))}~\delta q_{(BR (a))}  \\ \nonumber
&=& 2 \int_{{\cH}_{(BH)}} \delta T_{\mu}{}{}^{\nu} \xi^{\mu}_{(BH)}~ k_{\alpha (BH)} +
2 \sum_{a}~\int_{{\cH}_{(BR(a))}} \delta T_{\mu}{}{}^{\nu} \xi^{\mu}_{(BR (a))}~k_{\alpha (BR_{(a)})}
\een
The right-hand side of Eq.(\ref{rh}),
may be found by
the same procedure as described in Refs.\cite{gao01,rog02,rog05}. Namely,
considering $n$-dimensional Raychauduri Eq. and 
using the fact that
the null generators of the event horizon of the perturbed black ring coincide with
the null generators of the unperturbed stationary black ring, lead to the conclusion that
\be
\kappa_{(i)}~ \delta A_{(i)} = \int_{{\cal H}_{(i)}} \delta
T^{\mu}{}{}_{\nu}(matter) \xi^{\nu}_{(i)} k_{\mu (i)},
\label{ar}
\ee
where $\kappa_{(i)}$ is the surface gravity of the adequate black object.\\
In the light of what has been shown, according to Eqs.(\ref{rh}) and (\ref{ar}) we arrived at the
desired expression for
{\it physical process} version of the first law of black Saturn
mechanics in Einstein gravity with
additional
$(p+1)$-form field strength and dilaton fields. It yields
\ben
\alpha~ 
\bigg( 
\delta M_{BH} &+& \sum_{a} \delta {M_{BR}}^ {(a)}
 \bigg)
-  \sum_{i} \Omega_{(i)} \delta J^{(i)}_{BH} 
- \sum_{a}~ \sum_{i} \omega_{(i) (a)} \delta {J^{(i)}_{BR}}_{(a)} \\ \nonumber
&+&
\Phi_{(BH)}~\delta Q_{(BH)} + \sum_{a} \Phi_{(BR (a))}~\delta q_{(BR (a))} 
 = 2 \kappa_{(BH)} ~\delta {\cal A}_{(BH)} + 
2 \sum_{a}~\kappa_{(BR (a))} ~\delta {\cal A}_{(BR (a))}.
\een
One should have in mind that a proof of {\it physical process} version of the first law of
thermodynamics for $n$-dimensional black Saturn also provides support for cosmic censorship \cite{gao01}.  

%%%%%%%%%%%%%%%%%%%%%%%%%%%%%%%%%%%%%%%%%%%%%%%%%%%%%%%%%%%%%%%%%%%%%%%
%%%%%%%%%%%%%%%%%%%%%%%%%%%%%%%%%%%%%%%%%%%%%%%%%%%%%%%%%%%%%%%%%%%%%%%%%%%%%%%%%%%%%%%%%%%%%%%%%%%%%%%%%
\section{Equilibrium state version of the first law of black Saturn mechanics}
Now,
we would like to extend the above analysis and to
pay more attention to the first law of black Saturn dynamics 
by choosing an arbitrary cross section of the adequate event horizon of each black object to the future
of the bifurcation sphere. In Ref.\cite{gao03} it was shown that this attitude helped one to 
treat fields which were not necessarily smooth through the event horizon, having in mind 
the only requirement that the pull-back
of these fields in the future of the bifurcation surface be smooth. 
To begin with, let us consider
the spacetime with asymptotic conditions at infinity and equipped with the Killing vector fields
$\xi_{\mu (i)}$, which introduces an asymptotic symmetry. It 
was revealed in Ref.\cite{wal00} that there exist conserved quantities
$H_{\xi (i)}$, which imply
\be
\delta H_{\xi (i)} = \int_{\infty} \bigg( \bdel Q(\xi_{(i)}) - \xi_{(i)} \Theta \bigg).
\label{qua}
\ee
$\bdel$ is the variation which has no effect on $\xi_{\alpha}$ because of the fact that the Killing
vector field is treated as a fixed background and it ought not to be varied in expression (\ref{qua}).
In our case in Eq.(\ref{qua}) the symbol $(i)$ stands for black hole or each of for each of the black rings surrounded it.\\
In our considerations we were bound to the case of stationary axisymmetric $n$-dimensional black 
hole and $a$ black rings surrounded the black hole.
The 
Killing vector fields will be given by Eqs.(\ref{kil}) and (\ref{kilbr}).
In order to derive {\it equilibrium state} version of the first law of black Saturn mechanics
let us consider asymptotically hypersurfaces 
$\Sigma_{(i)}$ ending on the part of the event horizons ${\cal H}_{(i)}$ 
of each black object building black Saturn
to the future of the bifurcation surfaces. 
The cross sections of the black object horizons will constitute the inner boundaries of the hypersurfaces
$\Sigma_{(i)}$. It will be denoted by $S_{{\cal H} (i)}$, where $i$
stands respectively for black hole or $a$-th black ring. In our considerations 
we shall compare variations between two neighbouring states of the considered black objects.
There is a freedom which points can be chosen to correspond when one compares two
slightly different solutions. In our case we choose this freedom \cite{bar73}
to make $S_{{\cal H} (i)}$ the same of the two solutions (freedom of the general coordinate transformation)
as well as we consider the case when the null vector remains normal to $S_{{\cal H} (i)}$. Of course,
the stationarity and axisymmetricity of the solution will be preserved which in turn
causes that $\delta t^{\alpha}$, $\delta \psi^{\mu (i)}$ and $\delta \phi^{\mu (i)}$ will be equal to zero.
It provides that the variation of the Killing vector field $\xi_{\alpha (i)}$ is of the form, respectively for black hole
$\delta \xi^{\mu}_{(BH)} = \sum_{i} \delta \Omega_{(i)} \psi^{\mu (i)}$
and for $a$-th black ring $\delta \xi^{\mu}_{(BR (a))} = \sum_{i} \delta \omega_{(i) (a)} \phi^{\mu (i)}$  .
\par
Let us assume further that 
$(g_{\mu \nu}, B_{\alpha_{1} \dots \alpha_{p}}, \varphi)$ are solutions 
of the equations of motion and their variations
$(\delta g_{\mu \nu},~\delta B^{\alpha_{1} \dots \alpha_{p}},~\delta \varphi)$
are the linearized perturbations satisfying Eqs. of motion. 
Furthermore, one requires that the pull-back
of $B_{\alpha_{1} \dots \alpha_{p}}$ to the future of the bifurcation surface be smooth, but not 
necessarily smooth on it \cite{gao03}. Moreover, we require that $B_{\alpha_{1} \dots \alpha_{p}}$
and $\delta B_{\alpha_{1} \dots \alpha_{p}}$ fall off sufficiently rapid at infinity.
Then, those fields do not contribute to the canonical energy and canonical momenta.
For our black object we have the following relation:
\ben
\alpha 
\bigg(
\delta M_{BH} &+& \sum_{a} \delta {M_{BR}}^ {(a)}
 \bigg)
-  \sum_{i} \Omega_{(i)} \delta J^{(i)}_{BH} 
- \sum_{a}~ \sum_{i} \omega_{(i) (a)} \delta J^{(i)}_{BR (a)} \\ \nonumber
&=&
\int_{S_{\cal H} (BH)} \bigg( \bdel Q(\xi_{(BH)}) - \xi_{(BH)} \Theta \bigg) +
\sum_{a} \int_{S_{\cal H} (BR (a))} \bigg( \bdel Q(\xi_{(BR (a))}) - \xi_{(BR (a))} \Theta \bigg).
\een
First, we calculate the integral over the symplectic $(n-1)$-form bounded with the dilaton field.
We express $\ep_{\mu a j_{1} \dots j_{n-2}}$ by the volume element on $S_{{\cal H} (i)}$ and
by the vector $N^{\alpha}_{(i)}$, the {\it ingoing} future directed null normal to $S_{{\cal H} (i)}$, which is normalized
to $N^{\alpha}_{(i)} \xi_{\alpha (i)} = - 1$.
It gives the expression for each black object written as 
\ben
\int_{S_{{\cal H} (i)}}\xi^{j_{1}}_{(i)}~ \Theta_{j_{1} \dots j_{n-1}}^{\varphi} =
\int_{S_{{\cal H} (i)}} \ep_{j_{1} \dots j_{n-2}} N_{\alpha (i)} \xi^{\alpha}_{(i)}~ \xi_{\mu (i)} \na^{\mu} \varphi~
\delta \varphi = 0,
\een
where we used the fact that $\cL_{\xi} \varphi = 0$.\\
The same arguments as quoted in the previous section lead us to the following:
\be
\int_{S_{{\cal H} (BH)}} Q_{j_{1} \dots j_{n-2}}^{B}(\xi_{(BH)}) 
+ \sum_{a} \int_{S_{\cal H} (BR (a))} Q_{j_{1} \dots j_{n-2}}^{B}(\xi_{(BR (a))}) 
= \Phi_{(BH)}~Q_{(BH)} + \sum_{a} \Phi_{(BR (a))}~q_{(BR (a))}. 
\ee
The variation $\bdel$ of $Q_{j_{1} \dots j_{n-2}}^{B}(\xi)$ implies
\ben \label{bar}
\bdel \int_{S_{\cal H} (BH)} Q_{j_{1} \dots j_{n-2}}^{B} (\xi_{(BH)}) &+&
\sum_{a} \bdel \int_{S_{\cal H} (BR (a))} Q_{j_{1} \dots j_{n-2}}^{B} (\xi_{(BR (a))}) 
\\ \nonumber
&=& \delta (\Phi_{(BH)}~Q_{(BH)}) + \sum_{a} \delta (\Phi_{(BR (a))}~q_{(BR (a))}) \\ \nonumber
&-& {p \over (p + 1)!}\int_{S_{\cal H} (BH)} \sum_{i} \delta \Omega_{(i)} \psi^{\mu (i)}
B_{\mu \alpha_{3} \dots \alpha_{p+1}} 
\ep_{m k  j_{1} \dots j_{n-2}}~ e^{-{\alpha} \varphi}~ 
H^{m k \alpha_{3} \dots \alpha_{p+1}} \\ \nonumber
&-& {p \over (p + 1)!}
\sum_{a}
\int_{S_{\cal H} (BR (a))} \sum_{i} \delta \omega_{(i) (a)} \psi^{\mu (i)}_{(a)}
B_{\mu \alpha_{3} \dots \alpha_{p+1}} 
\ep_{m d j_{1} \dots j_{n-2}}~ e^{-{\alpha} \varphi}~ 
H^{m d \alpha_{3} \dots \alpha_{p+1}}.
\een
It can be easily verify that this relation and Eq.(\ref{bar}) enables us to write
\ben \label{pt}
\delta \Phi_{(BH)}~Q_{(BH)} 
&+& \sum_{a} \delta \Phi_{(BR (a))}~q_{(BR (a))} \\ \nonumber
&=& {p \over (p + 1)!}\int_{S_{\cal H} (BH)} \sum_{i} \delta \Omega_{(i)} \psi^{\mu (i)}
B_{\mu \alpha_{3} \dots \alpha_{p+1}} 
\ep_{m j j_{1} \dots j_{n-2}}~ e^{-{\alpha} \varphi}~ 
H^{m j \alpha_{3} \dots \alpha_{p+1}} \\ \nonumber
&+& {p \over (p + 1)!}
\int_{S_{\cal H} (BH)} \xi^{d}_{(BH)}~\delta B_{d \alpha_{3} \dots \alpha_{p+1}}
~N_{m}~\xi_{j (BH)}~e^{-{\alpha} \varphi}~ 
H^{m j \alpha_{3} \dots \alpha_{p+1}} \\ \nonumber
&+& {p \over (p + 1)!}
\sum_{a}
\int_{S_{\cal H} (BH)} \sum_{i} \delta \omega_{(i) (a)} \phi^{\mu (i)}_{(a)}
B_{\mu \alpha_{3} \dots \alpha_{p+1}} 
\ep_{m k j_{1} \dots j_{n-2}}~ e^{-{\alpha} \varphi}~ 
H^{m k \alpha_{3} \dots \alpha_{p+1}} \\ \nonumber
&+& {p \over (p + 1)!}
\sum_{a}
\int_{S_{\cal H} (BR (a))} \xi^{d}_{(BR) (a)}~\delta B_{d \alpha_{3} \dots \alpha_{p+1}}
~N_{m}~\xi_{k (BR) (a)}~e^{-{\alpha} \varphi}~ 
H^{m k \alpha_{3} \dots \alpha_{p+1}}
\een
Taking into account symplectic $(n-1)$-form for the potential
$B_{\nu_{1} \dots \nu_{p}}$ and the fact that on each event horizon of black object one has 
$H_{\mu \mu_{2} \dots \mu_{p+1}} \xi^{\mu}_{(i)} \sim \xi_{\mu_{2} (i)} \dots \xi_{\mu_{p+1} (i)}$
and expressing $\ep_{\mu a j_{1} \dots j_{n-2} (i)}$ in the same form
as in the above case, one arrives at the following:
\ben \label{bb1}
\int_{S_{\cal H} (BH)} \xi^{j_{1}}_{(BH)}~\Theta_{j_{1} \dots j_{n-1}}^{B} 
&+&
\sum_{a}
\int_{S_{\cal H} (BR (a))} \xi^{j_{1}}_{(BR (a))}~\Theta_{j_{1} \dots j_{n-1}}^{B}  \\ \nonumber
&=& {p \over (p+1)!}
\int_{S_{\cal H} (BH)} \ep_{j_{1} \dots j_{n-2}}~ e^{-{\alpha} \varphi}~ 
\xi_{k (BH)}~ H^{\delta k \nu_{3} \dots \nu_{p+1}}~N_{\delta (i)}~
\xi^{\nu_{2}}_{(BH)}~ \delta B_{\nu_{2} \dots \nu_{p+1}} \\ \nonumber
&+&
\sum_{a}
{p \over (p+1)!}
\int_{S_{\cal H} (BR (a))} \ep_{j_{1} \dots j_{n-2}}~ e^{-{\alpha} \varphi}~ 
\xi_{k (BR (a))}~ H^{\delta k \nu_{3} \dots \nu_{p+1}}~N_{\delta (i)}~
\xi^{\nu_{2}}_{(BR (a))}~ \delta B_{\nu_{2} \dots \nu_{p+1}}.
\een
According to Eq.(\ref{pt}) and (\ref{bb1}) 
one can conclude that the following is fulfilled:
\ben \label{char}
\bdel \int_{S_{\cal H} (BH)} Q_{j_{1} \dots j_{n-2}}^{B} (\xi_{(BH)})
&-& \xi^{j_{1}}_{(BH)}~\Theta_{j_{1} \dots j_{n-1}}^{B} 
+ \sum_{a}
\bdel \int_{S_{\cal H} (BR (a))} Q_{j_{1} \dots j_{n-2}}^{B} (\xi_{(BR (a))})
- \xi^{j_{1}}_{(BR (a))}~\Theta_{j_{1} \dots j_{n-1}}^{B} \\ \nonumber
&=& 
\Phi_{(BH)}~\delta Q_{(BH)} 
+ \sum_{a} \Phi_{(BR (a))}~\delta Q_{(BR (a))}. 
\een
For each black object one has
\be
\int_{S_{\cal H} (i)} Q_{j_{1} \dots j_{n-2}}^{GR} (\xi_{(i)}) = 2 \kappa_{(i)} A_{(i)},
\ee
where $A_{(i)} = \int_{S_{\cal H} (i)} \ep_{j_{1} \dots j_{n-2}}$ 
is the area of the black object horizon. It now follows that:  
\ben
\bdel \int_{S_{\cal H} (BH)} Q_{j_{1} \dots j_{n-2}}^{GR} (\xi_{(BH)}) 
&+& \sum_{a}
\bdel \int_{S_{\cal H} (BR (a))} Q_{j_{1} \dots j_{n-2}}^{GR} (\xi_{(BR (a))}) \\ \nonumber
&=&
2 \delta \bigg( \kappa_{(BH)} A_{(BH)} \bigg) +
2 \sum_{a} \bigg( \kappa_{(BR (a))} A_{(BR (a))} \bigg) \\ \nonumber
&+& 2 \sum_{i} \delta \Omega_{(i)}~ J^{(i)}_{(BH)} +
2 \sum_{a} \sum_{i} \delta \omega_{(i) (a)}~ J^{(i)}_{(BR (a))},
\een
where $J^{(i)}_{(BH)}= {1 \over 2}\int_{S_{\cal H} (BH)}\ep_{j_{1} \dots j_{n-2} a b} \na^{a} \psi^{(i)b}$
and  $J^{(i)}_{(BR (a))}= {1 \over 2}\int_{S_{\cal H} (BR (a))}\ep_{j_{1} \dots j_{n-2} a b} \na^{a} \phi^{(i)b}$
is the angular momentum connected with the Killing vector
fields responsible for the rotations in the adequate directions.
Having in mind calculations conducted in Ref.\cite{bar73} it could be verified that
the following integral is satisfied:
\ben
\int_{S_{\cal H} (BH)} \xi^{j_{1}}_{(BH)}~ \Theta_{j_{1} \dots j_{n-2}}^{GR} (\xi_{(BH)}) 
&+&
\sum_{a}
\int_{S_{\cal H} (BR (a))} \xi^{j_{1}}_{(BR (a))}~ \Theta_{j_{1} \dots j_{n-2}}^{GR} (\xi_{(BR (a))}) \\ \nonumber
&=&
2 A_{(BH)}~ \delta \kappa_{(BH)} + 2 \sum_{i} \delta \Omega_{(i)}~ J^{(i)}_{(BH)} \\ \nonumber
&+& 2 \sum_{a}
A_{(BR (a))}~ \delta \kappa_{(BR (a))} + 2 \sum_{a} \sum_{i} \delta \omega_{(i) (a)}~ J^{(i)}_{(BR(a))}.
\een                       
The above yields the conclusion that
\ben \label{arr}
\bdel \int_{S_{\cal H} (BH)} Q_{j_{1} \dots j_{n-2}}^{GR} (\xi_{(BH)})
&-& \xi^{j_{1}}_{(BH)}~\Theta_{j_{1} \dots j_{n-1}}^{GR} \\ \nonumber
&+& \sum_{a} \bigg[
\bdel \int_{S_{\cal H} (BR (a))} Q_{j_{1} \dots j_{n-2}}^{GR} (\xi_{(BR (a))})
- \xi^{j_{1}}_{(BR (a))}~\Theta_{j_{1} \dots j_{n-1}}^{GR} \bigg] \\ \nonumber
&=& 
2 \kappa_{(BH)}~\delta A_{(BH)}
+ 2 \sum_{a} \kappa_{(BR (a))}~\delta A_{(BR (a))}.
\een
As a direct consequence of relations (\ref{char}) and (\ref{arr}),
we find that we have obtained the first law of black Saturn mechanics in Einstein
$n$-dimensional gravity with additional $(p+1)$-form field strength and dilaton
fields. It may be written as    
\ben
\alpha~ 
\bigg( 
\delta M_{BH} &+& \sum_{a} \delta {M_{BR}}^ {(a)}
 \bigg)
-  \sum_{i} \Omega_{(i)} \delta J^{(i)}_{BH} 
- \sum_{a}~ \sum_{i} \omega_{(i) (a)} \delta {J^{(i)}_{BR}}_{(a)} \\ \nonumber
&+&
\Phi_{(BH)}~\delta Q_{(BH)} + \sum_{a} \Phi_{(BR (a))}~\delta q_{(BR (a))} 
 = 2 \kappa_{(BH)} ~\delta {\cal A}_{(BH)} + 
2 \sum_{a}~\kappa_{(BR (a))} ~\delta {\cal A}_{(BR (a))}.
\een

%%%%%%%%%%%%%%%%%%%%%%%%%%%%%%%%%%%%%%%%%%%%%%%%%%%%%%%%%%%%%%%%%%%%%%%%%%%%%%%%%%%%%%%%%%%%%%%%
\section{Conclusions}
In our paper we considered the first law of black Saturn thermodynamics in a higher dimensional theory
being the generalization of five-dimensional one with three-form field strength and dilaton field 
which contains stationary black hole and black ring solutions. 
Black Saturn under consideration is composed with stationary
$n$-dimensional black hole and $a$ black rings surrounded
this black hole. We provide both {\it physical process} version and {\it equilibrium state} version
of the first law of black Saturn mechanics. During the {\it physical process} version we change
infinitesimally black object under considerations by throwing matter into them. Assuming that black Saturn 
will be not destroyed in the process of it we find the changes of the ADM masses, angular momenta
as well as areas of the considered black objects.
\par
Considering {\it equilibrium state} version of the first law of black Saturn dynamics 
we chose arbitrary cross sections of each black object event horizons to the future
of bifurcation surfaces, contrary to the previous derivations which are bounded to the considerations
of bifurcation surfaces as the boundaries of hypersurfaces extending to spatial infinity.
Our attitude enables one to treat fields which are not necessary smooth through each event horizon of the adequate
black object.

%%%%%%%%%%%%%%%%%%%%%%%%%%%%%%%%%%%%%%%%%%%%%%%%%%%%%%%%%%%%%%%%%%%%%%%%%%%%%%%%%%%%%%%%%%%%%

%%%%%%%%%%%%%%%%%%%%%%%%%%%%%%%%%%%%%%%%%%%%%%%%%%%%%%%%%%%%%%%%%%%%%%%%%%%%%%%%%%%%%%%%%%%%%%%%
%\begin{appendix}

%\section{Irred   } 
%\label{irtf}
%\end{appendix}

%\begin{acknowledgments}
%MR was supported ny grant
%\end{acknowledgments}
%%%%%%%%%%%%%%%%%%%%%%%%%%%%%%%%%%%%%%%%%%%%%%%%%%%%%%%%%%%%%%%%%%%%%%%%%%%%%%%%%%%%%%%%%%
%%%%%%%%%%%%%%%%%%%%%%%%%%%%%%%%%%%%%%%%%%%%%%%%%%%%%%%%%%%%%%%%%%%%%%%%%%%%%%%%%%%%%%%%%%%%%%%%%%%%%%%
%%%%%%%%%%%%%%%%%%%%%%%%%%%%%%%%%%%%%%%%%%%%%%%%%%%%%%%%%%%%%%%%%%%
%%%%%%%%%%%%%%%%%%%%%%%%%%%%%%%%%%%%%%%%%%%%%%%%%%%%%%%%%%%%%%%%%%%%%%%%%%%%%%%%%
%%%%%%%%%%%%%%%%%%%%%%%%%%%%%%%%%%%%%%%%%%%%%%%%%%%%%%%%%%%%%%%%%%%%%%%%%%%%%%%%%

\end{document}